\documentclass[sigconf, authorversion, nonacm]{acmart}
\usepackage{algorithmic}
\usepackage{graphicx}
\usepackage{textcomp}
\usepackage{xcolor}
\usepackage[most]{tcolorbox}
\usepackage{printlen}
\usepackage{pgfplots}
\usepackage{float}
\usepackage{pgfplots}
\usepackage[font={small}]{caption}
\usepackage{multicol}
\usepackage{subcaption}
\PassOptionsToPackage{hyphens}{url}
\usepackage{cleveref}
\usepackage{listings}
\usepackage{tabularx}
\usepackage[htt]{hyphenat}
\usepackage{wrapfig}
\usepackage[export]{adjustbox}
\lstdefinestyle{yaml}{
     basicstyle=\color{blue}\footnotesize,
     rulecolor=\color{black},
     string=[s]{'}{'},
     stringstyle=\color{blue},
     comment=[l]{:},
     commentstyle=\color{black},
     morecomment=[l]{-}
 }

\newtcolorbox{myquote}[1][]{%
    colback=black!5,
    colframe=black!5,
    notitle,
    sharp corners,
    borderline west={2pt}{0pt}{black!80!black},
    enhanced,
    breakable,
    top=0.5pt,
    bottom=0.5pt
}

\graphicspath{{figures/}}

\def\BibTeX{{\rm B\kern-.05em{\sc i\kern-.025em b}\kern-.08em
    T\kern-.1667em\lower.7ex\hbox{E}\kern-.125emX}}

\usepackage{todonotes}

\usepackage{csquotes}

\pgfplotsset{compat=1.18}

\usepackage{array}


\begin{document}

\title{Analyzing the Evolution and Maintenance of ML Models on Hugging Face}


\author{Joel Castaño}
\affiliation{%
  \institution{Universitat Politècnica de Catalunya}
  \city{}
  \country{}}
\email{joel.castano@upc.edu}

\author{\normalfont Silverio Martínez-Fernández}
\affiliation{%
  \institution{
  Universitat Politècnica de Catalunya}
  \city{}
  \country{}
}
\email{silverio.martinez@upc.edu}

\author{Xavier Franch}
\affiliation{%
 \institution{Universitat Politècnica de Catalunya}
 \city{}
 \country{}
}
\email{xavier.franch@upc.edu}

\author{Justus Bogner}
\affiliation{%
  \institution{ Vrije Universiteit Amsterdam}
  \city{}
  \country{}
}
\email{j.bogner@vu.nl}

\renewcommand{\shortauthors}{Joel Castaño, Silverio Martínez, Xavier Franch, Justus Bogner}

\begin{abstract}
Hugging Face (HF) has established itself as a crucial platform for the development and sharing of machine learning (ML) models. This repository mining study, which delves into more than 380,000 models using data gathered via the HF Hub API, aims to explore the community engagement, evolution, and maintenance around models hosted on HF – aspects that have yet to be comprehensively explored in the literature. We first examine the overall growth and popularity of HF, uncovering trends in ML domains, framework usage, authors grouping and the evolution of tags and datasets used. Through text analysis of model card descriptions, we also seek to identify prevalent themes and insights within the developer community. Our investigation further extends to the maintenance aspects of models, where we evaluate the maintenance status of ML models,  classify commit messages into various categories (corrective, perfective, and adaptive), analyze the evolution across development stages of commits metrics and introduce a new classification system that estimates the maintenance status of models based on multiple attributes. This study aims to provide valuable insights about ML model maintenance and evolution that could inform future model development strategies on platforms like HF.
\end{abstract}



\maketitle

\section{Introduction}
The rapid evolution of machine learning (ML) models, especially on community platforms, is redefining the landscape of AI research and application. Hugging Face (HF) and its Hub~\cite{HuggingFaceInc.2023} stand out in this regard due to their critical role in the development, sharing, and deployment of a wide array of ML models, including Large Language Models (LLMs) and generative AI. HF represents an ecosystem where technical and social dynamics converge, forming a nexus of collaborative development that is continuously evolving. Despite its significance, the understanding of HF’s model evolution and maintenance practices remains underexplored.

Previous studies have explored various facets of HF, including pre-trained model reusability~\cite{Jiang2023}, the platform's carbon footprint~\cite{castano2023exploring},  or the challenges in reusing pre-trained models across different domains \cite{gong2023intended}. Our study aims to provide a holistic view of the current state of ML models on HF, focusing on their evolution, maintenance, and broader implications for the ML community. The novelty of our work lies in the comprehensive examination of these aspects on HF, which, to our knowledge, has not been explored in such detail before.

We delve into the dynamics of ML model maintenance and evolution on HF, investigating domain-specific trends, author collaboration patterns, content evolution in model cards, and detailed maintenance practices. These practices include analyses of commit types, file edits, and maintenance categorization, along with their correlations with model characteristics. Such information can guide users towards actively maintained models and inform their decision-making by highlighting the likelihood of future model updates. Moreover, our findings reveal the unique nature of ML model development compared to traditional software, emphasizing the diverse array of tools used, the crucial role of collaboration, and the distinctive developmental approaches.

The insights garnered from this study are not limited to the HF platform. They offer implications for the maintenance of ML models in general. The patterns and trends identified provide valuable lessons for the broader ML community, regardless of the specific platform, repository or environment used. Therefore, this research aims to guide the development of structured maintenance frameworks, enhancing transparency, and setting community-wide standards for ML model maintenance. Our study paves the way for future research in ML model evolution, offering both a framework and a replication package that can be adapted and applied beyond HF to improve maintenance activities of ML models in general.


\section{Background and Related Work}

\subsection{ML Model Maintenance and Evolution}
In recent decades, the rise in data and computing power availability has significantly enhanced ML applications across various domains \cite{Sarker2021}. ML models, integrated into ML systems \cite{Martinez-Fernandez2022}, require regular maintenance to address \textit{concept drift}—a decline in predictive accuracy over time due to changing data characteristics \cite{Lu2018, Leevy2020}. Maintenance tasks, crucial yet challenging \cite{Paleyes2023, Nazir2024}, involve ensuring operational stability and efficiency through corrections, adjustments, and optimizations, aligning with ISO 25059's software maintenance standards \cite{25010:2011} and interpretations by \citet{rowe1994defining}.

Contrastingly, evolution in ML models signifies substantial adaptations to new datasets and technologies \cite{Amershi2019}. As outlined by \citet{bennett2000software} and \citet{lehman1997metrics}, this encompasses a range of changes across the lifecycle of software, including the introduction of new features and meeting new requirements. Our research separates the overall evolution of the HF community, which involves trends in model development, framework usage, and author dynamics, from the specific maintenance of individual ML models that focuses on routine updates for current functionality.

\subsection{The Hugging Face Hub}
Training complex ML models requires considerable expertise and resources.
Therefore, it is advantageous to reuse existing pretrained models.
One community platform to facilitate such sharing and reuse is provided by the company Hugging Face, Inc. (HF).
Founded in 2016 as a Natural Language Processing (NLP) company, HF became popular for releasing their NLP models as open source~\cite{Jain2022} and creating a user-friendly library for NLP transformers~\cite{wolf2019huggingface}.
Today, the HF Hub represents their most important product, i.e., a public platform to train, share, download, and deploy ML models and datasets.
The Hub adopted the \textit{Model Cards} idea by \citet{mitchell2019model}: published models can provide a \texttt{README.md} and additional metadata, e.g., tags or prediction quality metrics, which leads to better documentation, transparency, and reproducibility. 
Models and datasets hosted on the Hub are represented as Git repositories\footnote{\url{https://huggingface.co/docs/hub/repositories}}, i.e., they are under version control, with multiple people being able to commit changes to them.
All in all, the HF Hub is slowly but surely establishing itself as the \enquote{GitHub of ML models}.\footnote{\url{https://www.forbes.com/sites/kenrickcai/2022/05/09/the-2-billion-emoji-hugging-face-wants-to-be-launchpad-for-a-machine-learning-revolution}}
However, unlike with GitHub, we know little about the state of the HF Hub and how the community uses it.

\subsection{Related Work}
Two types of publications are related to our study: a) publications about the maintenance and evolution of ML, and b) publications about analyzing the HF Hub.
Topic a) has mostly been studied from the perspective of maintainability challenges or technical debt, as visible through the secondary studies by, e.g., \citet{Shivashankar2022} and \citet{Bogner2021}.
However, some studies also analyzed maintenance and evolution activities in more detail.
\citet{Tang2021} analyzed 26 open-source ML projects on GitHub and studied how refactoring took place in these repositories to remove technical debt items.
Based on their analysis, they conceptualized new refactorings and technical debt categories specific to ML. 
They also proposed refactoring-related best practices and antipatterns.
\citet{Dilhara2021} conducted a mixed-method study to analyze the usage and evolution of ML libraries.
They first analyzed over 3,000 open-source repositories containing ML libraries and how their usage evolved.
Afterward, they surveyed 109 developers using ML libraries.
They identified that ML library updates frequently lead to the update of additional libraries, and that ML libraries are also downgraded in 20\% of the cases.
Lastly, they highlighted specific challenges for the maintenance and evolution of ML software.
To combat the decay of ML models through concept drift, \citet{Leest2023} proposed an architectural framework to support making design decisions to prepare an ML-enabled system for evolution.
The framework uses scenarios to capture different facets of evolution and to analyze trade-offs between evolvability and other quality attributes.
However, the framework has not been empirically evaluated so far, which the authors plan to do via industrial case studies.

Several studies also analyzed different characteristics of the HF Hub.
Taking a security perspective, \citet{Kathikar2023} analyzed the linked GitHub repositories of 110,000 HF models.
They used static analysis to identify a substantial number of vulnerabilities, even though the vast majority were of low severity.
However, the share of high-severity vulnerabilities was larger in popular fundamental repositories such as \texttt{Transformers}, which makes securing ML models even more complex.
In a previous study of ours \cite{castano2023exploring}, where we analyzed around 170,000 models to uncover insights about HF’s impact on environmental sustainability, we discovered that only a very tiny percentage of models reported the carbon emissions from their training. Most of these were models trained on the HF infrastructure, which reports these emissions automatically. Over the years, the share of models reporting carbon emissions also decreased, but for those that did report them, mean emissions decreased slightly. We also identified factors correlating with high carbon emissions.
\citet{Ait2023} wanted to make the analysis of HF more convenient and therefore created \textit{HFCommunity}, a tool that collects and integrates data about the HF Hub, e.g., data on repositories, discussions, files, commits, etc.
The data is provided as a relational database dump that can be downloaded and analyzed offline.
The authors envision \textit{HFCommunity} as a long-term data source to enable efficient empirical studies of ML projects.
\citet{Jiang2023} conducted an interview study with practitioners who use HF.
They identified practices and challenges regarding the reuse of pretrained models.
Afterward, they extended their data with a security risk analysis based on information mined from the HF Hub.
They concluded that several risky practices exist in the supply chain of pretrained models, e.g., a frequent lack of signatures.
Lastly, \citet{gong2023intended} explored pre-trained model usage across repositories like HF. They emphasized the need for \enquote{model contracts} to address challenges in reusing models due to domain gaps, recommending specifications on intended usage, limitations, and performance for better model reuse.

While several studies have analyzed the maintenance and evolution of ML software, no study reports about these activities for models on the HF Hub.
Getting insights into how ML models are maintained and evolve in the largest community platform could lead to the identification of important challenges and practices, and can also inform more design-oriented future research.

\section{Methodology}
In this section, we outline our methodology, stating from the study objective and research questions, followed by an explanation of the dataset collection process. 


\subsection{Study Objective and Research Questions}

Following the Goal Question Metric (GQM) guidelines \cite{caldiera1994goal}, our research goal is structured as follows:\\
Analyze \textit{pre-trained ML models} for the purpose of \textit{exploring and categorizing} with respect to \textit{their present status, evolution and  maintenance} from the viewpoint of \textit{ML researchers and practitioners} in the context of \textit{the HF Hub}.

Two main research questions (RQ) arise from this goal. We explore the models in HF to understand their development, popularity, and maintenance:

\begin{myquote}
\textbf{RQ1}. \textit{What is the current status and evolution of the HF community?}
\end{myquote}

\begin{itemize}
    \item RQ1.1: How has HF's popularity changed?
    \item RQ1.2: How have framework usage, tag, and dataset trends evolved in HF?
    \item RQ1.3: Are there prominent authors groups in HF community?
    \item RQ1.4: What trends and insights can be identified from the content of HF model cards?
\end{itemize}

\begin{myquote}
\textbf{RQ2}. \textit{How can we evaluate and categorize the maintenance status of ML models on HF through their commit information?}
\end{myquote}

\begin{itemize}
    \item RQ2.1: What do commit metrics reveal about the maintenance of ML models?
    \item RQ2.2 How does the size and frequency of commits evolve over time?
    \item RQ2.3: How do different types of commits (perfective, corrective, adaptive) contribute to the maintenance of models? 
    \item RQ2.4 How do the editing patterns of specific files evolve across different development stages?
    \item RQ2.5: How can we classify the maintenance status of individual models using their commit data?
    \item RQ2.6: How do various model characteristics differ between maintenance levels?
\end{itemize}

\subsection{Dataset Construction}

To answer our RQs, we execute a data collection and preprocessing pipeline, refined to meet the specific demands and objectives of the current study. The data extraction process was carried out on November 6th, 2023.

\textbf{Data availability statement}: The datasets, code, and detailed documentation are available in a replication package hosted on Zenodo \cite{anonymous_2023_10153155}.

\subsubsection{\textbf{Data Collection}}

Our data collection pipeline employs the HF Hub API using the HfApi class \cite{huggingfaceHfApiClient}, a Python wrapper, to collect data about users and models stored on the HF platform. To this end, we collect a range of common model attributes, including: the total size of datasets used, hardware used for training, evaluation metrics such as accuracy or F1, size of the model file in the repository, number of downloads and likes for each model, tags attached to each model (e.g., PyTorch, Transformer), the raw text of the model's card and more. For more detailed information on the data attributes and collection process, refer to \cite{anonymous_2023_10153155}.

In addition to these common attributes, our pipeline is enhanced to collect detailed data related to the commit history of models, providing insights into their development and maintenance over time. This approach is complemented by the integration of data from the \textit{HFCommunity} dataset \cite{Ait2023}, an offline up-to-date relational database built from the data available at the HF Hub. The \textit{HFCommunity} dataset used the PyDriller framework to extract detailed commit information, thereby providing access to the list of files edited in each commit, a feature not available through the HF API. This additional layer of data enriches our analysis by offering a more complete view of the changes made to each model over time.

In addition to commit data, we retrieve discussion data, which includes questions, pull requests, and issues related to the models from the HF API. More details on the data collection are deferred to the replication package.

\subsubsection{\textbf{Data Preprocessing}}

Our analysis involves the processing of the newly incorporated commit and discussion data. The dataset after the collection phase possesses over 380,000 data entries, each representing a model on HF. 

Firstly, we classify commit data to assess the nature of changes made to the models based on its messages. This classification aligns with Swanson's traditional software maintenance categories — Corrective, Perfective, and Adaptive \cite{swanson1976dimensions}. The classification of commits is performed using a neural network approach based on the work of \citet{sarwar2020multi}, who fine-tuned an off-the-shelf neural network, DistilBERT, for the commit message classification task. We fine-tune the neural network proposed in the paper and use it to classify each of the commits.

Beyond classification, we derive metrics that reflect model evolution and ongoing maintenance efforts, such as the frequency and distribution of various commit types. This process is critical for understanding the lifecycle and robustness of the models.

Moreover, we harmonize variables, manage missing values and identify and handle irrelevant or low-impact attributes appropriately ensuring the dataset's integrity and consistency. The final step in our preprocessing is the application of one-hot encoding to the tags associated with the models. This encoding, combined with a developed tag-to-domain dictionary allows us to filter and map tags to domains, including: Multimodal, Computer Vision, NLP, Audio, and Reinforcement Learning.

\subsection{Data Analysis}

In this section, we describe the methodology for analyzing the data to answer our research questions. We aim to provide a clear and reproducible account of how we analyzed the data and derived conclusions.

\subsubsection{\textbf{RQ1 Analysis}}
To address RQ1.1, we construct several time-series graphs using attributes that could indicate and demonstrate an increase in popularity. Specifically, we analyze trends in the number of new models added each month, the number of commits created, likes, the number of new unique authors, and the number of opened discussions. 

For RQ1.2, we analyze the overall statistics of datasets, tags, and libraries, followed by a time-series plot that demonstrates the proportion of the attributes that have been in the top 5 each year. This analysis helps us to identify the trends and popularity of specific tags, datasets, and libraries over time. To assess whether there are statistically significant evolutionary differences over time in the usage of top frameworks (\textit{pytorch}, \textit{tensorflow}, and \textit{jax}), we employ a Chi-squared test for independence. This test is particularly suitable for our analysis, as it allows us to evaluate the relationship between categorical variables (in this case, the frameworks) across multiple time periods.

To address RQ1.3, we employ a graph-based approach to uncover groups of authors using the Louvain algorithm \cite{blondel2008fast}. We choose the Louvain method for its efficiency and effectiveness detecting communities in a large network such as HF. We construct a graph with authors as nodes and collaborations as edges, attributing model popularity to each author and linking co-authors. We define the popularity on each author as the sum of the popularity of each model they collaborated on. The Louvain method was then employed for community detection, identifying clusters of closely connected authors (i.e., author groups). Subsequently, we calculated the cumulative popularity of each author group, ranked them in descending order, and visualized the concentration of popularity among the top groups.

Lastly, for RQ1.4, we use Latent Dirichlet Allocation (LDA) \cite{blei2003latent} to identify common topics and their evolution within the text of the model cards. For LDA's hyperparameters tuning, we conduct experiments with topic coherence metrics, such as $C_v$ \cite{roder2015exploring}, and perform manual inspection of the topics to ensure they are distinct and meaningful. This approach is complemented by testing various hyperparameters, including document and word topic densities.

\subsubsection{\textbf{RQ2 Analysis}}
For the maintenance status of models, we analyze the descriptive statistics of five main maintenance metrics (RQ2.1): the number of commits per model, average number of files edited by the commits for a model, monthly commit frequency, and the total number of authors involved in a commit, as in HF a single commit for a model can be made by multiple authors.

For RQ2.2, we examine the evolutionary trend of the number of commits and commit size, employing a slope t-test on a fitted linear regression with a significance level of $\alpha=0.05$ to test for any significant trends.

For RQ2.3, we use the classified maintenance types (perfective, corrective, and adaptive) to analyze their proportions, and how they evolve throughout a model's development lifecycle. That is, we calculate the proportion of commit types across development stages (from beginning to end) and plot the evolution of these proportions throughout the development process.

For RQ2.4, we identify the most commonly edited files in commits and analyze the file editing lifecycle to observe how the patterns of these edits change over the course of a model's development. Equivalently with RQ2.3, we measure the proportion of edits to specific files at five key stages in the development cycle, providing insights into the evolving nature of file modifications as the model matures. Finally, to better understand the relationships between files commonly edited together, we construct a graph where nodes represent individual files, and edges denote the co-occurrence of file edits within the same commit. The weight of each edge corresponds to the frequency of these co-editing events, offering a quantitative measure of the strength of the relationship between files. Using again the Louvain algorithm, we then detect communities within this graph. These communities are groups of files that are frequently edited together, which we describe and visualize.

For RQ2.5, we employ a k-means clustering algorithm to classify the maintenance status of ML models on the HF platform. We opt for $k=2$ based on initial observations of minimal variance for $k>2$, ensuring a distinction between high and low maintenance models, aligning with similar approaches such as \citet{coelho2020github}. K-means is selected for its simplicity and effectiveness in generating distinct, interpretable clusters, ideal for delineating clear maintenance groups. In contrast, methods like DBSCAN \cite{schubert2017dbscan}, which automatically determines the number of clusters based on data density, might not explicitly align with our specific objective of categorizing into two maintenance categories. The classification is based on several key features: the total number of commits, the frequency of commits per month, the average interval between commits, the longest duration without any commits, the count of contributing authors, and the proportion of discussions that have been successfully closed. The selection of these features was influenced by their relevance to maintenance activities, and was also informed by similar attributes used in parallel research \cite{coelho2020github}. Details can be further seen in the replication package.

Finally, for RQ2.6, we investigate how various model characteristics differ between high and low maintenance categories. Our analysis encompasses both continuous and nominal variables, using statistical tests appropriate for non-normally distributed data at significance level $\alpha=0.05$. For continuous variables, including likes, downloads, size, model card text length, accuracy, f1, and datasets size, we employ the Mann-Whitney U test \cite{mcknight2010mann}. This test is suitable for comparing the means of two independent samples without assuming normality. For nominal variables, specifically the domain and library usage, we use the Z-test for proportions. This test allows us to determine whether the proportion of models in a particular domain or using a specific library significantly differs between the high and low maintenance categories.

\section{Results}

In this section, we present the results of the data analysis per RQ.

\subsection{Current Status and Evolution of HF Community (RQ1)}
~
\subsubsection{\textbf{How has HF's popularity changed?}}
\begin{figure}[h]
\centering
\includegraphics[width=0.7\columnwidth]{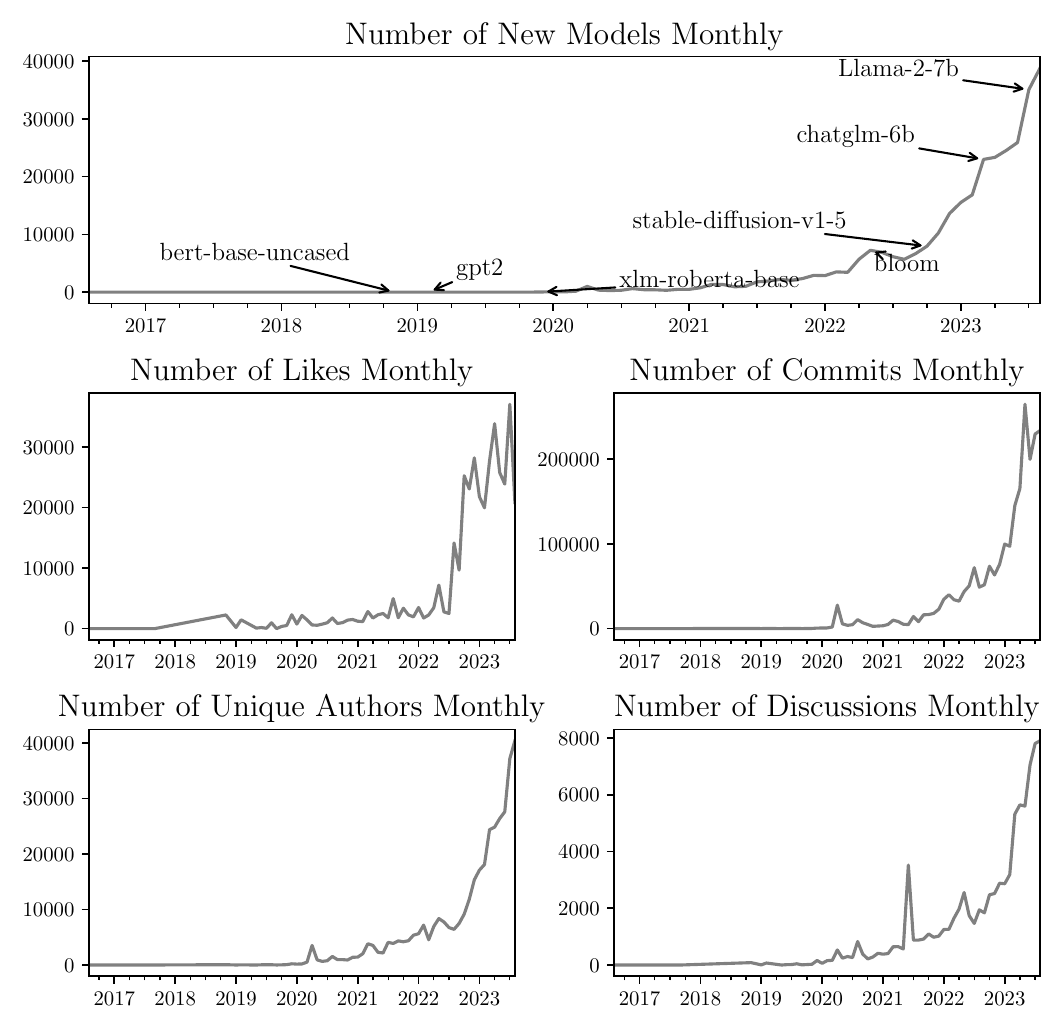}
\caption{Popularity metrics evolution on HF.}
\label{popularity_evolution}
\end{figure}

~
The results displayed in Figure~\ref{popularity_evolution} clearly showcase HF's exponential growth in popularity.

The analysis of various metrics from 2017 to 2023 on the HF platform reveals a consistent and significant uptrend in engagement and development activities. Notably, there has been a clear increase in the number of new models added each month. This trend is accompanied by a marked spike in community engagement, as evidenced by the rise in likes and discussions, particularly from 2022 onwards. Additionally, there has been a steady growth in both the number of commits and the diversity of contributing authors each month. These trends collectively highlight the growing importance and popularity of HF in the ML community, aligning with findings from previous studies ~\cite{castano2023exploring}.

\begin{myquote}
\textbf{Finding 1.1}. \textit{HF's popularity has exponentially increased over time, which is evident from the upward trends in the number of new models, likes, commits, unique authors, and discussions aggregated monthly.}
\end{myquote}

\subsubsection{\textbf{How have framework usage, tags and datasets trends evolved in HF?}}
Regarding the most common libraries, \textit{transformers} and \textit{pytorch} hold the first and second positions respectively, with \textit{transformers} being used in 163,936 models and \textit{pytorch} in 150,757 models. This is further supported by the top frameworks on each domain, where \textit{transformers} and \textit{pytorch} appear across domains such as Audio, Computer Vision, and NLP. For Multimodal models, the top library is \textit{diffusers}, which has state-of-the-art pretrained diffusion models for generating images and audio. For Reinforcement Learning tasks, we have \textit{Stable-Baselines3}, a set of implementations of reinforcement learning algorithms built on top of PyTorch.

\begin{figure}[h]
\centering

\begin{subfigure}{\columnwidth}
  \centering
  \includegraphics[width=0.7\linewidth]{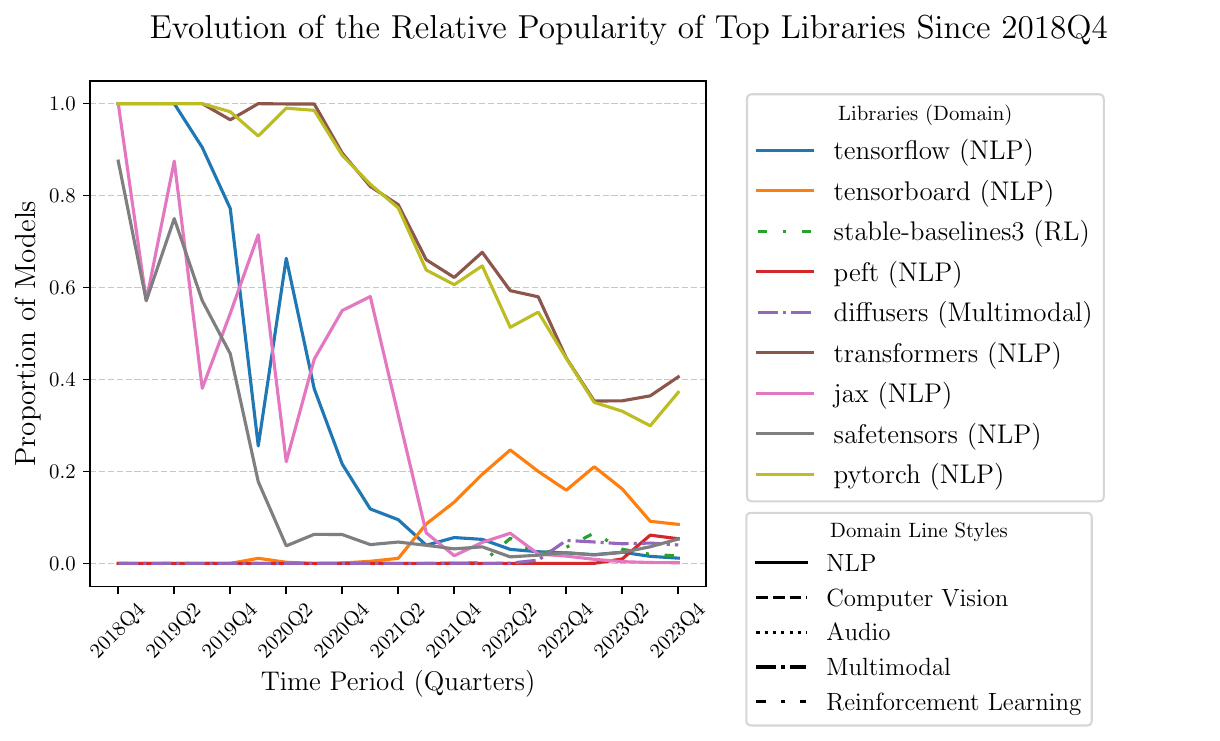}
  \label{fig:libraries_evolution}
\end{subfigure}

\begin{subfigure}{\columnwidth}
  \centering
  \includegraphics[width=0.7\linewidth]{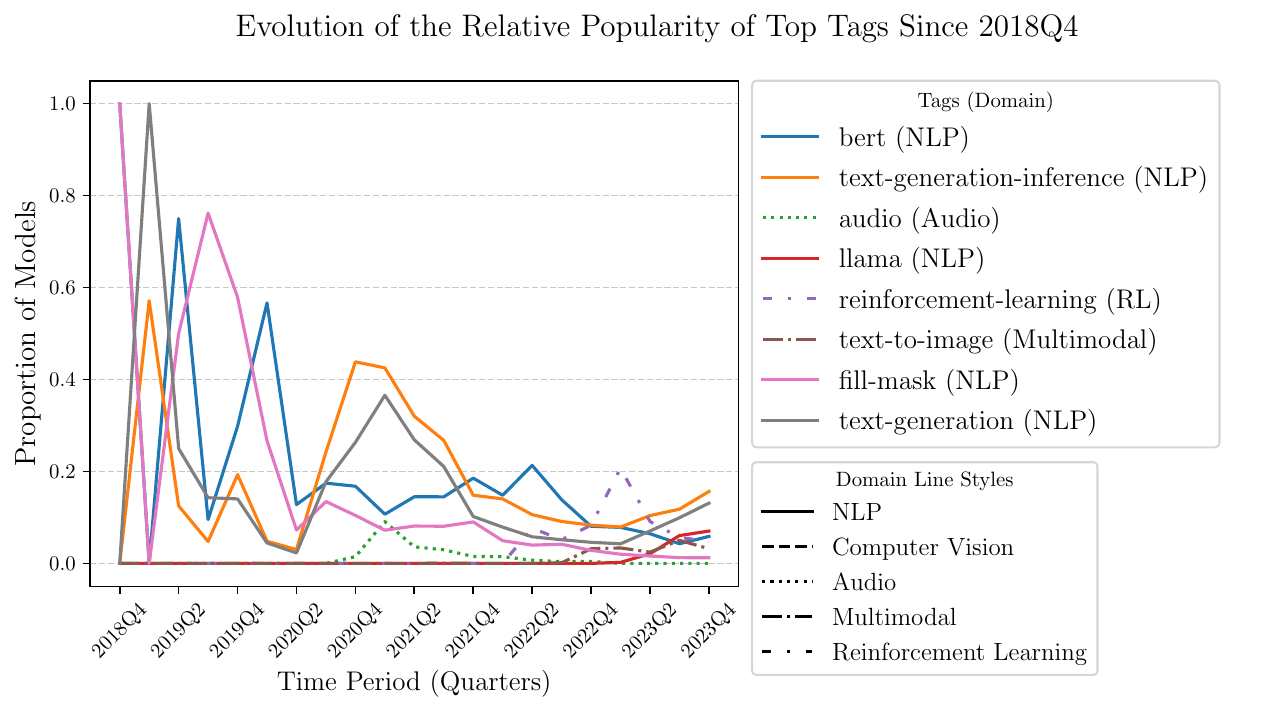}
  \label{fig:evolution_top_tags}
\end{subfigure}

\begin{subfigure}{\columnwidth}
  \centering
  \includegraphics[width=0.7\linewidth]{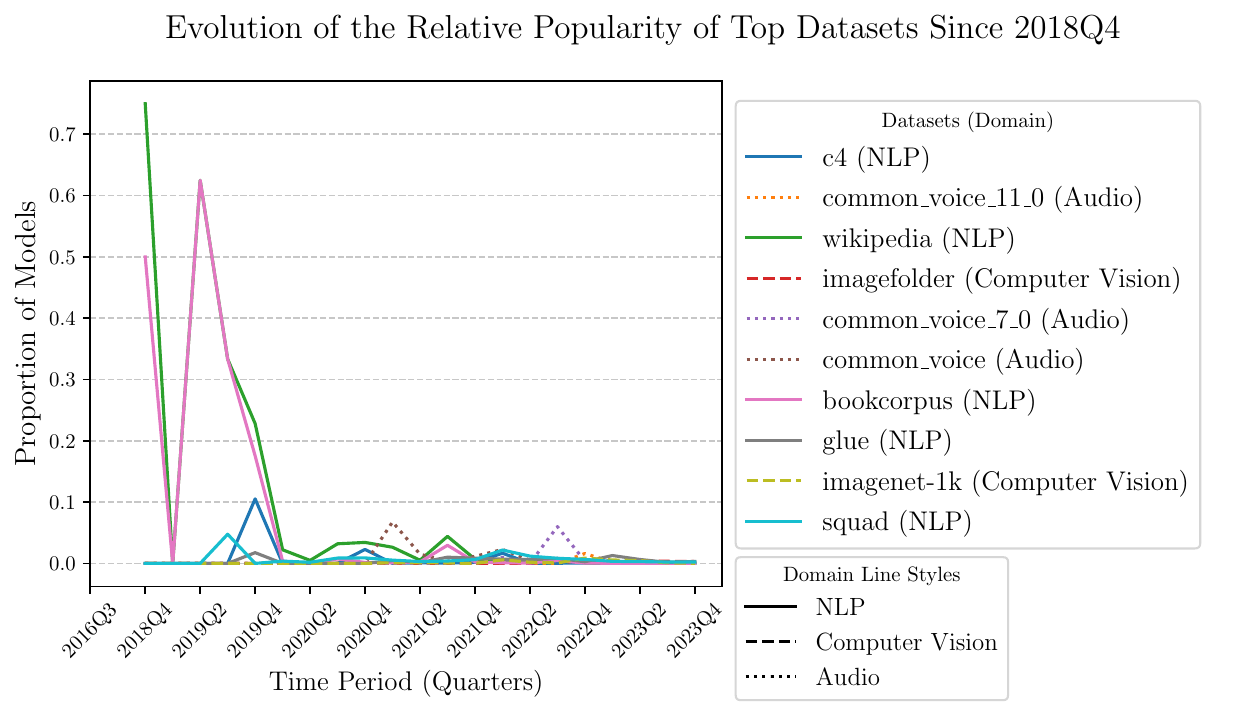}
  \label{fig:evolution_top_datasets}
\end{subfigure}
\caption{Evolution of the relative popularity}

\label{fig:library_tags_dataset_evolution}
\vspace{-0.2cm}
\end{figure}

In total, there are 129 unique libraries used across various models on HF. The evolution of the relative popularity of top libraries can be seen in Figure~\ref{fig:library_tags_dataset_evolution}.
This figure is a line chart showing the trends of  the proportion of various libraries from the fourth quarter of 2018 to the fourth quarter of 2023. It is important to note that a single model can incorporate multiple libraries, resulting in several libraries having high proportions (e.g., \textit{tensorflow}, \textit{jax} and \textit{safetensors} at 2019Q1). Furthermore, we recognize that certain libraries like the \textit{transformers} (HF \textit{Transformers} library \cite{wolf2020transformers}) may be used in conjunction with others such as \textit{pytorch} or \textit{tensorflow}. In our analysis, these are only counted if they are explicitly mentioned to prevent false positives. The following observations can be made:
\begin{itemize}
\item \textbf{Dominant Libraries}: \textit{pytorch} and \textit{transformers}   are consistently the most dominant tags, although the total proportion of models with these tags has been shrinking until reaching $\approx$ 40\% today, suggesting an increased variety of tags.  The observed decline reflects the increasing diversity of models on HF. As the platform grew, a wider array of models emerged, leading to a dilution in the proportion of models using these libraries.
\item \textbf{Reinforcement Learning Libraries}: \textit{stable-baselines3} experienced a surge in popularity at the beginning of 2022, but it remains less prominent compared to major NLP libraries.
\item \textbf{Multimodal Libraries}: \textit{diffusers} experienced a surge in popularity in mid-2022, ranking as the fifth most popular library in the last quarter.
\item \textbf{Library Comparison}: In comparing popular frameworks, \textit{pytorch} remains the most popular, while \textit{tensorflow} and \textit{jax} have seen notable decreases in usage. Specifically, \textit{pytorch}'s usage declined by 62.79\%, whereas \textit{tensorflow} and \textit{jax} experienced sharper declines of 98.85\% and 99.85\%, respectively. The Chi-square test confirms these trends are statistically significant (p-value<0.05).
\end{itemize}

\begin{myquote}
\textbf{Finding 1.2.1}. \textit{`transformers' and `pytorch' are the most used frameworks overall. Moreover, `pytorch' maintained the framework dominance while `tensorflow' and `jax' lost popularity.}
\end{myquote}

As we delve further into the evolution of HF, we observe a shift in the popularity of tags and datasets over the years. As for the tags, there are a total of 23,496 unique tags on HF. The most frequent tags encompass a range of topics, from library-specific and auxiliary tags to language-related tags. However, when we filter out these non-specific tags, we uncover the true interests of the HF community. Generative AI and NLP-related tags such as \textit{text-generation-inference}, \textit{text-classification}, and \textit{reinforcement-learning} are particularly prevalent.

Figure~\ref{fig:library_tags_dataset_evolution} also illustrates the dynamic landscape of tag popularity over the years. We observe a steady decline in older NLP models like BERT, while tags related to generative AI have gained momentum. Although there is a consistent interest in audio-related tasks and reinforcement learning, they remain less popular than some of the NLP tags. The last three quarters have witnessed a surge in interest for \texttt{text-generation-inference}, \texttt{text-to-image}, and \texttt{llama}.

\begin{myquote}
\textbf{Finding 1.2.2} \textit{The analysis of tags reveals a dominant interest in generative AI and NLP within the HF community, with notable but less significant interest in other domain tags such as audio-related tasks and reinforcement learning.}
\end{myquote}


Finally, in terms of datasets, a total of 10,525 unique datasets are used. In Figure~\ref{fig:library_tags_dataset_evolution}, we can also observe the changing popularity of datasets over time. In the earlier years, NLP datasets like GLUE and Wikipedia were foundational, serving as benchmark datasets. However, their popularity has waned over time, possibly due to the emergence of newer datasets or shifting research priorities, without new hugely dominant datasets on the community.

\begin{myquote}
\textbf{Finding 1.2.3}. \textit{Although NLP datasets like GLUE and Wikipedia were foundational in earlier years, there are no dominant datasets nowadays.}
\end{myquote}

\subsubsection{\textbf{Who are the prominent authors and what are their relationships in HF?}}

An analysis of the most common authors reveals names such as 'SFconvertbot' and 'librarian-bot', which are automated bots contributing to the platform among other human authors.

\begin{center}
\begin{tabularx}{\columnwidth}{*{2}{>{\centering\arraybackslash}X}}
    \centering
    \includegraphics[width=0.4\columnwidth]{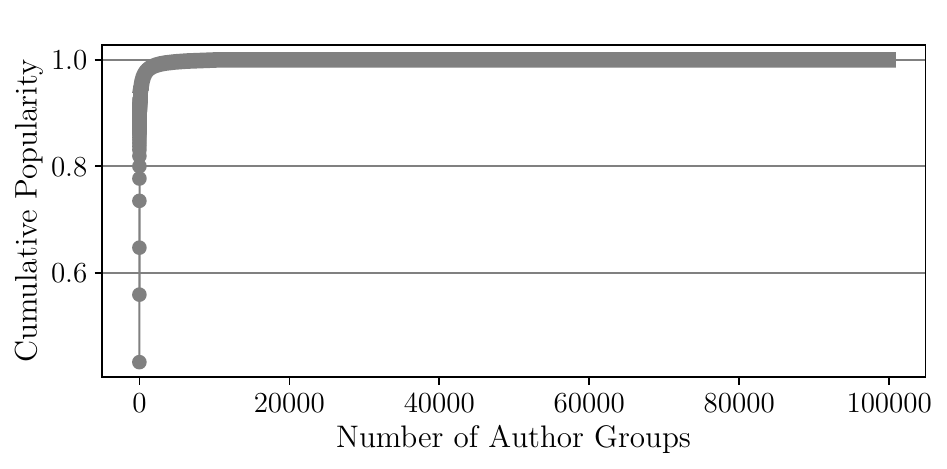}
    \captionof{figure}{Cumulative popularity of author groups}
    \label{cumulative_popularity_author_groups}
& 
    \vspace{-1.9cm}
    \centering
    \footnotesize
    \begin{tabular}{ll}
        \textbf{Metric}                                    & \textbf{Value} \\
                    \hline
                    \hline
        Average Downloads                                  & 1,858,571   \\
        Average Likes                                      & 294.71         \\
        Average \# of Authors                          & 8.61           \\
        Average Length of Card                       & 8,807.55       \\
                                              \\
    \end{tabular}
    \captionof{table}{Average Statistics Top Author Group}
    \label{table:average_model_statistics}
    
\end{tabularx}
\vspace{-0.85cm}
\end{center}

To further explore the relationships between authors, we employed the Louvain algorithm to identify clusters of authors who frequently collaborate with each other. The results were quite telling. A small number of groups garnered the majority of popularity on HF, with the first group alone accounting for 40\% of the platform's popularity as can be seen in Figure~\ref{cumulative_popularity_author_groups}. This is further evidenced in Table ~\ref{table:average_model_statistics}, which displays statistics for the top author group models that significantly surpass the average metrics of a typical HF model. For example, while the average number of likes for a model is 1.13, the top group boasts an average of 294.71. This group consisted of approximately 580 authors.

The dominance of this group is further emphasized when considering that HF has over 100,000 unique authors. This means that a tiny fraction of authors (approximately 0.5\%) are responsible for a significant portion of the platform's popularity. The collaborative nature of these authors is evident in their contributions to models with extensive collaboration such as \texttt{bigscience/bloom} with 22 unique collaborators, among others.

These findings reveal a clear concentration of popularity among a small number of authors who tend to collaborate frequently, indicating a tight-knit community of contributors who play a significant role in shaping the landscape of HF.

\begin{myquote}
\textbf{Finding 1.3}. \textit{A small number of author groups, particularly one dominant group, convey the majority of popularity in HF. This indicates a concentrated popularity among authors who often collaborate with each other.}
\end{myquote}

\subsubsection{\textbf{What trends and insights can be identified from the content of model cards?}}

An LDA decomposition was conducted to categorize the prevalent themes within the model card content on all published model cards on HF (87,775). We chose symmetric Dirichlet priors, assigning equal prior weight to each topic, as this yielded similar results across multiple choices and ensured a balanced representation of topics. With five components, we ensured that the topics generated were distinct and meaningful. The identified raw topics (top 10 words for each topic) were:

\begin{itemize}
    \item \textbf{Topic 1: Training Info}
        \begin{itemize}
            \item \textit{Raw}: "training information needed model loss hyperparameters evaluation results following"
            \item \textit{Interp.}: About model training specifics.
        \end{itemize}
    \item \textbf{Topic 2: Text Generation}
        \begin{itemize}
            \item \textit{Raw}: "model huggingface llama 7b information use needed 13b prompt models"
            \item \textit{Interp.}: Linked to text generation, with references to llama, prompting or common number of parameters.
        \end{itemize}
    \item \textbf{Topic 3: Reinforcement Learning}
        \begin{itemize}
            \item \textit{Raw}: "agent model td playing baselines3 false python stable github rl"
            \item \textit{Interp.}: Models centered on agent-based learning.
        \end{itemize}
    \item \textbf{Topic 4: NSFW Content}
        \begin{itemize}
            \item \textit{Raw}: "png previews click nsfw style f1 strong font suit maid"
            \item \textit{Interp.}: Generation of explicit adult content via HF's NSFW mainly with text-to-image generative content.
        \end{itemize}
    \item \textbf{Topic 5: Other}
        \begin{itemize}
            \item \textit{Raw}: "model huggingface main github import image resolve use trained models"
            \item \textit{Interp.}: Includes miscellaneous and uncategorized cards.
        \end{itemize}
\end{itemize}


\begin{figure}[h]
\centering
\includegraphics[width=0.5\columnwidth]{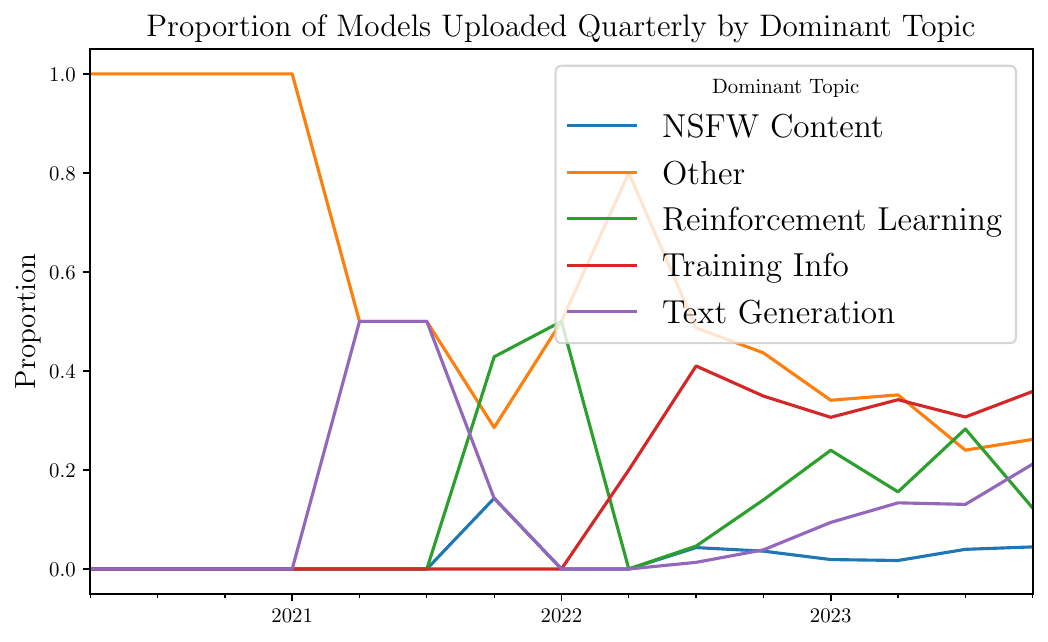}
\caption{Time series analysis of model cards LDA topics}
\vspace{-0.2cm}
\label{topic_evolution}
\end{figure}

As observed, our topics do not align perfectly with the categories presented by \citet{mitchell2019model}, suggesting unique trends and focuses in the HF community to explore further.

The evolution of these topics is presented in Figure~\ref{topic_evolution}. A growing popularity of NSFW content since the mid 2022 aligns with the urge in popularity of image diffusion models. Simultaneously, the consistent rise in text generation, also propelled by the generative AI wave, emphasizes its lasting significance.
"Training Information" dominates the model card topics, indicating that model-specific training details remain in model cards.
Lastly, the cyclical nature of "Reinforcement Learning", echoing similar patterns in tag evolution, highlights periodic surges in interest, potentially aligned with advancements or novel applications in this field.

\begin{myquote}
\textbf{Finding 1.4}. \textit{Model cards combine technical terms, training parameters, and general descriptors, indicative of the nature their content. Emerging trends like generative AI underscore evolving user interests and applications.}
\end{myquote}

\subsection{Maintenance Analysis (RQ2)}

\subsubsection{\textbf{What do commit patterns and classifications reveal about the diversity in model development on HF?}}

\begin{figure}[h]
\centering
\includegraphics[width=0.7\columnwidth]{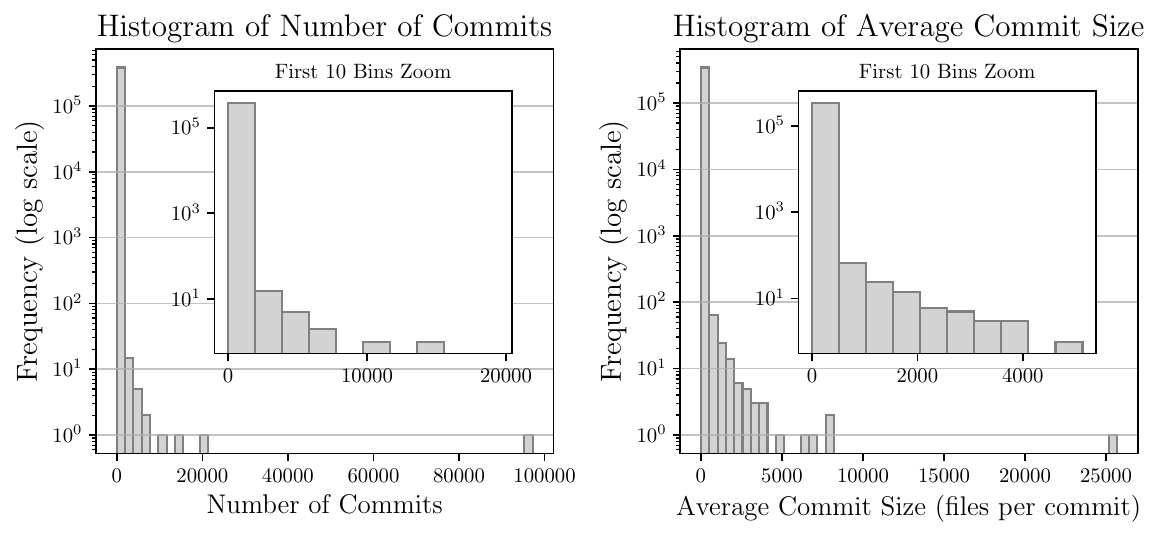}
\caption{\# of commits and commit size per model histograms}
\label{commits_histograms}
\end{figure}

The diversity in development activities among HF models is clearly evidenced by the range in the number of commits (Figure~\ref{commits_histograms}). The average number of commits is 7.16, but the median is only 3.0, showing a strong skew in the distribution with a mean substantially higher than the median. This discrepancy is also reflected in the top models by the number of commits. A few models, such as \texttt{CivitAI\_model\_info} with 97,237 commits, have undergone extensive revisions and updates. In contrast, a significant number of models have very few commits (e.g., 71\% of the models have less than 5 commits), suggesting disparities in active development or maintenance. The abnormally large numbers of commits on some models are mostly attributed to automatic commits using the HF API, e.g., updating the model file with a commit on every training step, thereby giving an illusion of high maintenance.

\begin{myquote}
\textbf{Finding 2.1.1}. \textit{HF is characterized by a diverse but right-skewed distribution of commit patterns (influenced by automated processes), with few models receiving extensive updates and the majority showing limited activity.}
\end{myquote}

The commit size and frequency further highlights the diversity in development practices (Figure~\ref{commits_histograms}). While the average commit size is 5.0 files per commit, the median is 2.0, indicating that most changes are incremental and minor, but there are occasional substantial modifications that could represent significant feature additions or improvements to a model. The frequency of commits also exhibit a wide range, with some models being updated frequently, while others have long intervals between commits: the mean time between commits is 52.6 days, while the median is 27 minutes.

\begin{myquote}
\textbf{Finding 2.1.2}. \textit{Incremental improvements dominate model development, as evidenced by the prevalence of minor commits.}
\end{myquote}

When we look at the 100 most popular models, we can identify models with high maintenance metric values from actual development processes rather than automated commits. For instance, the top models by number of commits are \texttt{OrangeMixs} with 185 commits, \texttt{bloom} with 108 commits, and \texttt{chatglm-6b} with 95 commits. Similarly, models such as \texttt{lllyasviel/ControlNet-v1-1} with an average commit size of 5.66 files per commit showcase substantial modifications indicative of significant developmental efforts.

The involvement of diverse contributors in model development on HF ranges from individual developers to collaborative efforts. While the majority of models are developed by a small number of authors (with a median of 1 author per model), there are noteworthy exceptions. Models like \texttt{bigscience/bloom} or \texttt{bigcode/santacoder} are examples of collaborations that bring 22 and 17 unique authors respectively.

\begin{myquote}
\textbf{Finding 2.1.3}. \textit{HF encompasses both individual and collaborative efforts. While the average number of unique authors per model is low, (1.18 mean and 1.0 median), there are notable examples of collaboration.}
\end{myquote}

\subsubsection{\textbf{How does the size and frequency of commits evolve over time?}}

\begin{figure}[h]
\centering
\includegraphics[width=0.7\columnwidth]{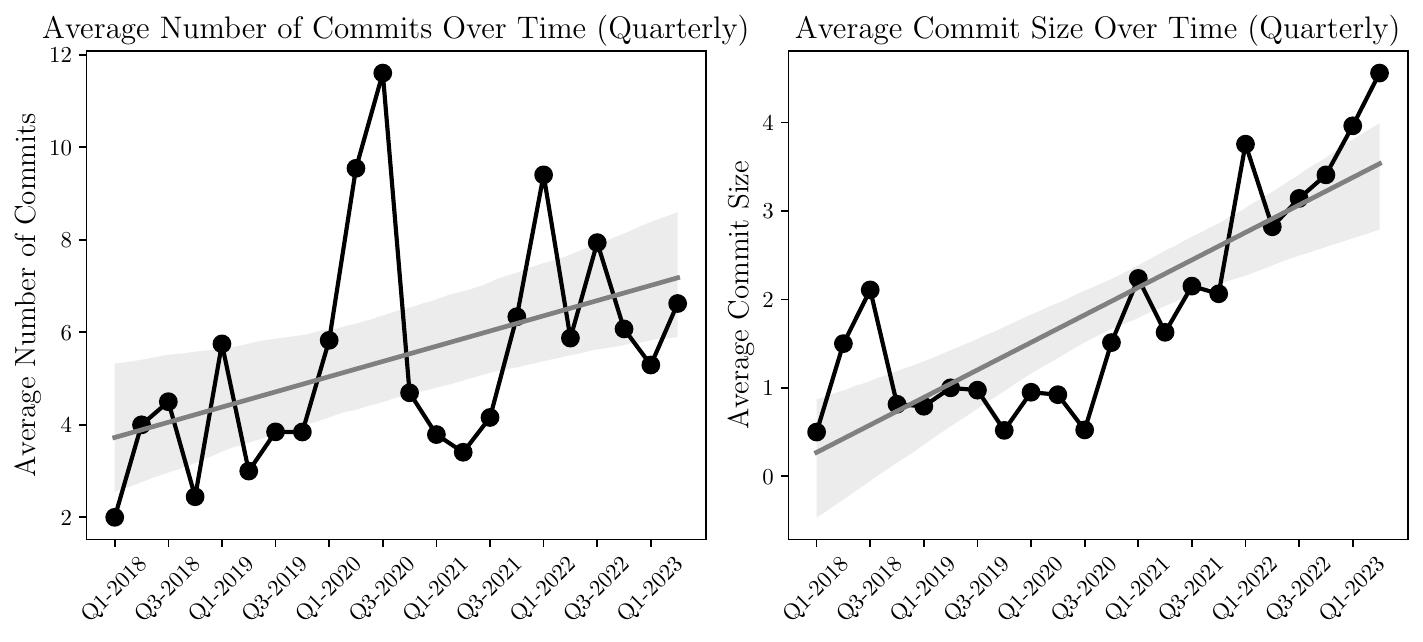}
\caption{Average number of commits for each model quarterly }
\label{commits_and_size_quarterly}
\end{figure}

Figure~\ref{commits_and_size_quarterly} presents the quarterly evolution of the number of commits and the average commit size for all models.

As observed in Figure~\ref{commits_and_size_quarterly}, there is a slight upward trend in both the average number of commits and the average commit size over time. This suggests that there might be an increase in the maintenance efforts put into the models hosted on the HF platform. This increase can be attributed the increasing awareness of the importance of model maintenance, an overall increase in the quality and complexity of the hosted models, or a more distributed usage of the HF API, which makes automatic commits easier. The p-value on the slope t-test on both trends is  $<0.05$, confirming that this is not just a random fluctuation, but a significant statistical trend. Furthermore, the peak observed in the number of commits and the corresponding decrease in the average commit size during 2020-Q3 can be attributed to the 'Helsinki-NLP' organization. They uploaded a substantial number of models (over 300), characterized by a high frequency of commits and a low number of files edited per commit.

\begin{myquote}
\textbf{Finding 2.2}. \textit{There has been a slight but statistically significant increase in the average number of commits and the average commit size over time, indicating a possible increase in model maintenance efforts on the HF platform.}
\end{myquote}

\subsubsection{\textbf{How do different types of commits (perfective, corrective, adaptive) contribute to the maintenance of models on the HF platform?}}

The classification algorithm \cite{sarwar2020multi} provided multiple labels, including combinations between perfective, corrective, and adaptive. We present examples for each commit type:

\begin{itemize}
    \item \textbf{Corrective Commits:}
    \begin{itemize}
        \item \texttt{'Updated bug in TensorFlow usage code \\(README.md) (\#5)'}
        \item \texttt{'[FIX] Fix Typo (\#3)'}
    \end{itemize}
    
    \item \textbf{Perfective Commits:}
    \begin{itemize}
        \item \texttt{'For clarity, delete deprecated modelcard.json'}
        \item \texttt{'Update tokenizer.json'}
    \end{itemize}
    
    \item \textbf{Adaptive Commits:}
    \begin{itemize}
        \item \texttt{'Adding `safetensors` variant of this model (\#1)'}
        \item \texttt{'Add size details'}
    \end{itemize}
\end{itemize}

As we can observe, corrective commits address bugs and errors, such as fixing typos or updating incorrect code. Perfective commits focus on improvements and refinements, like updating a tokenizer or deleting deprecated files. Lastly, adaptive commits add new features or variants.

By analyzing the commit type frequencies of 2,760,224 commits, we found a dominant proportion of perfective commits, as showcased in Table~\ref{tab:commit-types}.

Considering \textit{Finding 2.1.2}, it is reasonable to infer that the majority of the commits are perfective in nature. The high proportion of perfective commits aligns with the trend of incremental small improvements being dominant in model development.In fact, significant portion of perfective commits corresponds to routine updates such as `update README.md', `update pytorch\_model.bin', and other similar enhancements.

\begin{myquote}
\textbf{Finding 2.3.1}. \textit{Perfective commits constitute the majority of maintenance activities on HF. This suggests a strong emphasis on incremental improvements and routine updates during the model development lifecycle.}
\end{myquote}

\begin{center}
\begin{tabularx}{\columnwidth}{*{2}{>{\centering\arraybackslash}X}}
    \centering
    \includegraphics[width=0.5\columnwidth]{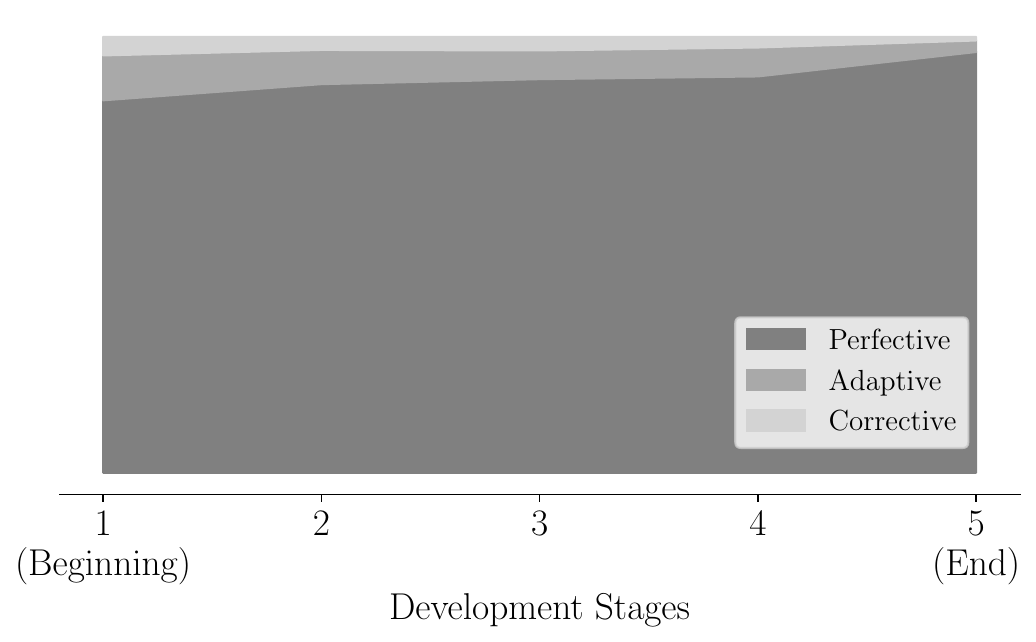}
    \captionof{figure}{Lifecycle of commit types}
    \label{fig:evolution_commit_types}
& 
    \vspace{-2.8cm}
    \centering
    \footnotesize
    \begin{tabular}{lr}
        \textbf{Commit Type} & \textbf{(\%)} \\
        \hline
        \hline
        Perfective & 89.3 \\
        Adaptive & 6.1 \\
        Corrective & 2.46 \\
        Adaptive Perfective & 1.85 \\
        Corrective Perfective & 0.08 \\
        Corrective Adaptive & $<$0.01 \\
        Unclassified & 0.15 \\
    \end{tabular}
    \vspace{0.2cm}
    \captionof{table}{Proportions of commit types}
    \label{tab:commit-types}
\end{tabularx}
\end{center}
\vspace{-0.6cm}

Further, we sought to understand the lifecycle of commit types over the development of an average model. The results, depicted in Figure~\ref{fig:evolution_commit_types}, which shows the proportion of different types of commits across various stages of development, indicates a noticeable trend.
From the graph, we observe the following:
\begin{itemize}
\item Perfective commit are dominant at the start and increase even more towards the end development stage.
\item Adaptive commits decrease as development progresses.
\item Corrective commits maintain a consistent, thin layer throughout the stages with a slight decrease at the end.
\end{itemize}

The chart suggests that model development on the HF platform typically begins with a mix of perfective and adaptive tasks, gradually transitioning towards perfective efforts, which dominate towards the end. This implies that, as the development matures, there are fewer environmental or requirement changes (adaptive) and a stable number of bug fixes or issue resolutions (corrective), with an increasing focus on enhancing existing features (perfective).

\begin{myquote}
\textbf{Finding 2.3.2}. \textit{Throughout the lifecycle of a model, there is an increase on perfective tasks. This indicates a maturing of model development where enhancements take precedence over new features or the rectification of defects.}
\end{myquote}

\subsubsection{\textbf{How do the editing patterns of specific files evolve across different development stages?}}

\begin{figure}[ht]
    \centering
    \begin{minipage}[b]{0.49\columnwidth}
        \centering
        \includegraphics[width=\columnwidth]{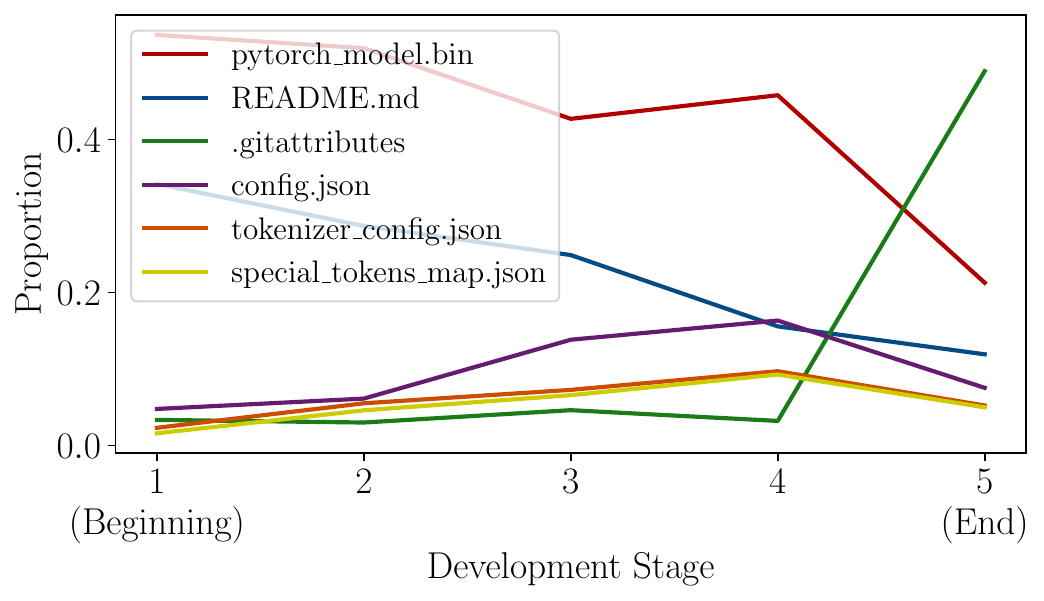}
        \caption{Lifecycle of files edited in commits}
        \label{evolutionary_analysis_filename_editing}
    \end{minipage}
    \hfill
    \begin{minipage}[b]{0.49\columnwidth}
        \centering
        \includegraphics[width=\columnwidth]{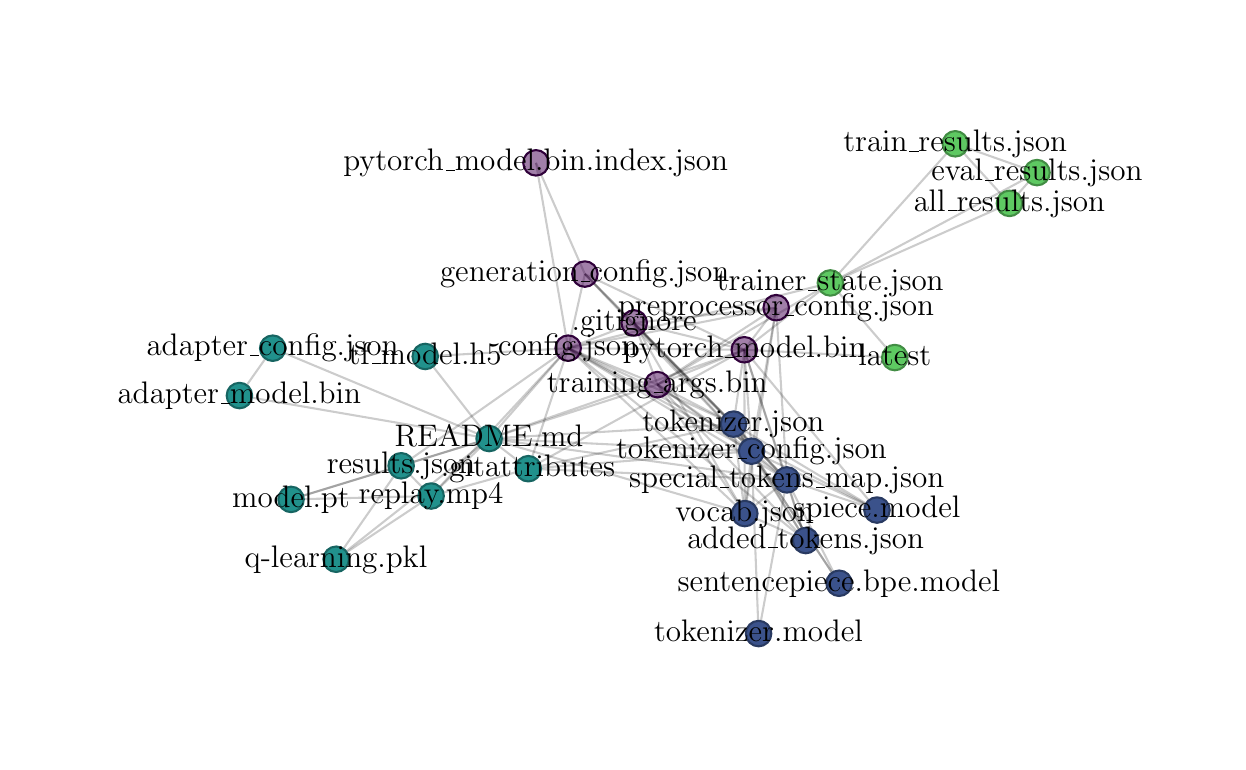}
        \caption{Files Network Graph Clustering}
        \label{files_clustering_graph}
    \end{minipage}
\end{figure}

An analysis of the most commonly edited files reveals \textit{pytorch\_model.bin} as the most edited filename followed by \textit{README.md} and \textit{.gitattributes}. The lifecycle of files edited in commits, depicted in Figure~\ref{evolutionary_analysis_filename_editing}, provides further insights.
From the figure, we can draw several conclusions:
\begin{itemize}
\item The editing proportion of \textit{pytorch\_model.bin} decreases over development stages, indicating its core component status and decreasing need for modifications.
\item A significant spike in edits for \textit{.gitattributes} at stage 5.
\item A slow and steady drop in edits for \textit{README.md}.
\item An increase in edits to \textit{config.json} during the middle stages of development, followed by a gradual decline.
\item Files such as \textit{special\_tokens\_map.json} and \textit{tokenizer\_config.json} experience a marginal number of edits throughout the entire development stage.
\end{itemize}

Therefore, the development lifecycle of the typical HF model is characterized by an initial phase of active development and fine-tuning, followed by a stabilization phase where fewer changes are needed. This reflects the iterative process of model development, where the initial focus is on getting the architecture and weights right, followed by optimization and fine-tuning, and finally documentation and other supporting files. Additionally, Figure \ref{files_clustering_graph} illustrates the clustering derived from the Louvain algorithm applied to a graph representing files that are commonly edited at the same time (for a higher resolution version, refer to the replication package). This analysis identified four primary clusters: tokenizer-related files (e.g., \textit{tokenizer.json}, \textit{tokenizer\_config.json}), model and configuration files (such as \textit{pytorch\_model.bin}, \textit{training\_args.bin}), training results data (including examples like \textit{train\_results.json}, \textit{eval\_results.json}), and a miscellaneous cluster featuring files such as \textit{README.md} and \textit{.gitattributes}. These clusters represent groupings of files that are frequently edited together at the same time.

\begin{myquote}
\textbf{Finding 2.4.1}. \textit{Reduced edits in \textit{pytorch\_model.bin} indicate a shift from initial development to stability, with changes in \textit{README.md} or \textit{config.json} marking the transition from setup to final tuning in model development.}
\end{myquote}

\begin{myquote}
\textbf{Finding 2.4.2}. \textit{The clustering analysis identifies file clusters that represent files frequently edited concurrently (tokenizer, model, training results, and miscellaneous files) highlighting synchronized editing patterns and interdependencies.}
\end{myquote}

\subsubsection{\textbf{Classification of Model Maintenance Using Commit Data}}

Our k-means clustering algorithm segregated the ML models into distinct maintenance categories based on their activity patterns in the repository, resulting in two primary categories:

\begin{itemize}
    \item \textbf{High Maintenance Category:} Models with active maintenance practices, characterized by a higher number of commits, regular commit frequency, shorter intervals between commits, fewer days without commits, and a slightly higher number of authors.
    \item \textbf{Low Maintenance Category:} Models with less frequent maintenance activities, indicated by fewer commits, lower frequency of commits, longer intervals between commits, more days without commits, and fewer authors involved.
\end{itemize}

With 62,818 models classified as high maintenance and 319,477 models as low maintenance (83.5\% vs 16.4\% respectively) the classification underscores the diverse nature of model maintenance within the HF ecosystem. An analysis of the centroids for both clusters provides quantitative insights into the maintenance behaviors:

\begin{table}[h!]
\setlength{\abovecaptionskip}{-5pt}
\setlength{\belowcaptionskip}{5pt}
\centering
\resizebox{\columnwidth}{!}{%
\begin{tabular}{|l|c|c|c|c|c|c|}
\hline
\textbf{Category} & \textbf{\begin{tabular}[c]{@{}c@{}}Num\\ Commits\end{tabular}} & \textbf{\begin{tabular}[c]{@{}c@{}}Commit\\ Frequency\end{tabular}} & \textbf{\begin{tabular}[c]{@{}c@{}}Avg Days\\ Between Commits\end{tabular}} & \textbf{\begin{tabular}[c]{@{}c@{}}Max Days\\ Without Commits\end{tabular}} & \textbf{\begin{tabular}[c]{@{}c@{}}Num\\ Authors\end{tabular}} & \textbf{\begin{tabular}[c]{@{}c@{}}\% Closed\\ Discussions\end{tabular}} \\ \hline
High Maintenance & 28.7 & 8.0 & 8.0 & 170.7 & 1.5 & 17.3\\ \hline
Low Maintenance & 3.0 & 0.6 & 61.2 & 255.9 & 1.1 & 0.1 \\ \hline
\end{tabular}%
}
\caption{Mean Centroids of Maintenance Categories}
\label{tab:maintenance_centroids}
\end{table}

\begin{figure}[h]
\centering
\includegraphics[width=0.7\columnwidth]{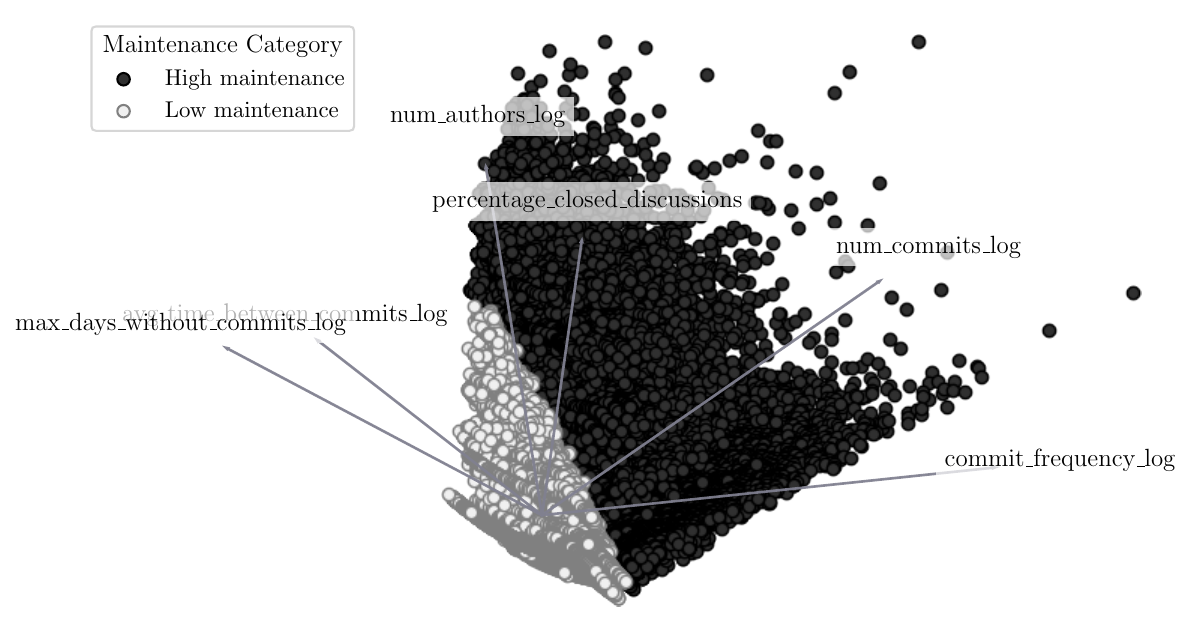}
\caption{Biplot of K-Means Clustering on Maintenance Features}
\label{fig:classification_clustering}
\end{figure}

The mean centroids clearly illustrate the distinction between high and low maintenance models, with high maintenance models showing greater engagement and activity. In Figure \ref{fig:classification_clustering}, the biplot derived from a Principal Component Analysis (PCA) visually represents the separation between these two maintenance categories. This figure effectively represents the division of models into high and low maintenance categories, with each dot representing a model, and the positioning informed by the maintenance attributes.

\begin{myquote}
\textbf{Finding 2.5}. \textit{K-means clustering categorized 16.4\% of HF models as 'High Maintenance' and 83.5\% as 'Low Maintenance', underscoring diverse maintenance practices and clear distinctions in engagement, as reflected in the centroids.}
\end{myquote}

\subsubsection{\textbf{How do various model characteristics differ between maintenance levels?}}

Our analysis revealed significant differences in several model characteristics between high and low maintenance categories. We summarize the findings in two tables: one for continuous variables analyzed using the Mann-Whitney U test and another for nominal variables using the Z-test for proportions.

\begin{table}[H]
\centering
\setlength{\abovecaptionskip}{-5pt}
\setlength{\belowcaptionskip}{5pt}
\resizebox{\columnwidth}{!}{%
\begin{tabular}{|l|c|c|c|}
\hline
\textbf{Variable} & \textbf{High Maintenance Mean} & \textbf{Low Maintenance Mean} & \textbf{p-value} \\ \hline
Popularity & 0.00073 & 0.000029 & <0.001 \\ \hline
Likes & 5.51 & 0.25 & <0.001 \\ \hline
Downloads & 11,390.46 & 241.04 & <0.001 \\ \hline
Size (MB) & 5,976,630.0 & 1,308,191.6 & <0.001 \\ \hline
Model Card Text Length & 4,132.9 & 2,070.9 & <0.001 \\ \hline
Accuracy & 0.8002 & 0.8202 & 0.13 \\ \hline
F1 Score & 0.754 & 0.7370 & 0.748 \\ \hline
Dataset Size (MB) & 48,672,083.7 & 33,105,320.8 & 0.12 \\ \hline
\end{tabular}
}
\caption{Mann-Whitney U Test Results for Continuous Variables}
\end{table}

\vspace{-0.5cm}

\begin{table}[H]
\centering
\setlength{\abovecaptionskip}{-5pt}
\setlength{\belowcaptionskip}{5pt}
\resizebox{\columnwidth}{!}{%
\begin{tabular}{|l|c|c|c|}
\hline
\textbf{Variable} & \textbf{High Maintenance Proportion} & \textbf{Low Maintenance Proportion} & \textbf{p-value} \\ \hline
NLP & 0.7303 & 0.6248 & <0.001 \\ \hline
Audio & 0.0987 & 0.0798 & <0.001 \\ \hline
Computer Vision & 0.0684 & 0.0499 & <0.001 \\ \hline
Multimodal & 0.0886 & 0.0670 & <0.001 \\ \hline
Reinforcement Learning & 0.0140 & 0.1784 & <0.001 \\ \hline
tf & 0.0552 & 0.0222 & <0.001 \\ \hline
jax & 0.0319 & 0.0208 & <0.001 \\ \hline
transformers & 0.6965 & 0.3830 & <0.001 \\ \hline
pytorch & 0.6473 & 0.3414 & <0.001 \\ \hline
\end{tabular}
}
\caption{Z-Test for Proportions Results for Nominal Variables}
\end{table}

The results indicate that models with high maintenance tend to be more popular, have more likes and downloads, and are larger in size compared to those with low maintenance. In fact, when we select the leading author group identified in RQ1.3 and classify the top 500 models that have the most authors in common with this group, we find that 98\% of these models fall into the high maintenance category. This suggests that the concentration of popular author groups also extends to a concentration of high maintenance activities within specific author groups. Additionally, the model card text length is significantly longer for high maintenance models, suggesting more extensive documentation in this category.

In terms of nominal variables, the domains of NLP, Audio, Computer Vision, and Multimodal showed a higher proportion for high maintenance, whereas Reinforcement Learning was more prevalent for low maintenance. For libraries, 'transformers' and 'pytorch' were predominantly used in high maintenance models, whereas the others showed significant but less pronounced differences.


\begin{myquote}
\textbf{Finding 2.6}. \textit{High-maintenance models tend to be more popular, larger, and better documented than their low-maintenance counterparts, with a notable concentration of high maintenance activities within specific author groups.}
\end{myquote}

\section{Implications}

This study offers a comprehensive examination of the evolution and maintenance practices within the HF community, providing significant insights that can spearhead advancements in the ML domain. These have relevant implications for both researchers and practitioners, providing them with a deeper understanding of the dynamics in model evolution that can inform best practices for model maintenance and evolution in community-driven platforms.

\subsection{Status and Evolution of Hugging Face (RQ1)}

The evolutionary insights presented significant trends in model development on HF, offering a predictive lens for the future trajectory of ML research and applications. Our findings chart the progressive details of model evolution, providing a valuable barometer for the ML community's direction.

\begin{itemize}
    \item \textbf{Predictive Trends for Strategic Alignment:} By mapping the growth patterns in model additions and framework usage (Finding 1.1 and 1.2), we provide a predictive foundation for researchers and developers to strategically align their efforts with future demands and community directions.
    \item \textbf{Emphasis on Collaboration:} The insights into authorship dynamics (Finding 1.3 and Finding 2.6) indicate that high-maintenance models, which are more popular and better documented, often emerge from these collaborative multi-author environments, emphasizing the impact of collective efforts on model quality and visibility.
    \item \textbf{Model Documentation as a Reflective Mirror:} The growing emphasis on generative AI in model cards, as seen in Finding 1.4, underscores the dynamic development of ML and the need for robust documentation. This evolving landscape underscores the necessity for robust documentation, echoing \citet{oreamuno2023state}'s observation of inadequate documentation in many HF models and datasets. This issue is compounded by Finding 2.5's revelation of prevalent low maintenance in models, aligning with \citet{bhat2023aspirations}'s call for responsible, up-to-date documentation practices. As ML models evolve in complexity, it is imperative that their documentation maintains high standards of clarity, completeness, and ethical considerations, enhancing accountability and usability across applications.

\end{itemize}

\subsection{Maintenance and Evolution of Models (RQ2)}

Our analysis reveals significant variance in maintenance practices across HF models, underscoring the need for systematic, collaborative, and continuously refined maintenance approaches.

\begin{itemize}

    \item \textbf{Understanding File Edit Interdependencies:} Finding 2.4.2 reveals how clustering analysis can highlight synchronized editing patterns and interdependencies among file types. This knowledge is valuable for anticipating and managing linked changes in model files, which can streamline maintenance processes and minimize errors. Furthermore, if HF facilitated the retrieval of line-level change data through its API, the analysis of file interdependencies could be greatly enhanced. Zimmermann et al. \cite{zimmermann2006msr} have demonstrated the importance of mining version archives at the line level. This granular approach could provide deeper insights into ML model evolution.

    \item \textbf{Lifecycle Planning:} Understanding the typical lifecycle of model development (Finding 2.4.1) offers practical benefits for developers in optimizing their maintenance strategies. For instance, the transition from frequent edits in \textit{pytorch\_model.bin} to adjustments in \textit{README.md} or \textit{config.json} can be used as indicators to recognize when a model is shifting from its development phase to stabilization and refinement. This awareness enables developers to allocate resources more efficiently, ensuring that the development resources are used at the right stages of the model's lifecycle, leading to more efficient and effective model evolution.
    
    \item \textbf{Refined Maintenance Categorization:} The implementation of the maintenance classification, as revealed in Finding 2.5, not only enables users to make more informed choices by identifying models that are actively maintained but also introduces the potential for a Long Term Support (LTS) model in HF. This approach would categorize certain models as LTS, indicating a commitment to longer-term stability, regular updates, and support, enhancing transparency and ensuring that users can rely on these models for extended periods without significant changes disrupting their projects. For developers, this structured framework provides a clear roadmap for prioritizing maintenance tasks.

\end{itemize}

\subsection{ML Systems vs. Traditional Repositories}

The maintenance of ML models on HF presents a unique pattern compared to the maintenance of traditional software systems. In the realm of traditional software system development, such as those found in repositories like GitHub, the focus typically lies on bug fixes and feature additions. This involves version releases and systematic testing cycles \cite{pressman2005software}, reflecting a development paradigm where changes are often driven by evolving user requirements or efforts at software optimization.

In HF's ML model development, we noted individual and collaborative efforts, highlighted by the varied nature of maintenance activities. This variance is reflected in the distribution of commit patterns, where perfective maintenance emerges as the dominant approach (Finding 2.3). Such an approach, focusing on incremental model improvements, contrasts with traditional software system development seen in repositories like GitHub. In the HF context, the maintenance of ML models prioritizes enhancing model performance and aligning with evolving technological advancements.

This trend indicates a departure from the traditional software maintenance paradigms. It reveals the need for methods and tools specifically designed for the unique demands of ML model maintenance. Such tools may include advanced version control systems optimized for data and model tracking, as well as automated monitoring tools capable of detecting model drift or degradation. These tools and methodologies should align with the principles of continuous learning, model monitoring, and dynamic adaptation to data changes, which are crucial for maintaining the quality and relevance of ML models over time. The exploration of MLOps tools such as DVC \cite{dvcDataVersion} and DagsHub \cite{dagshubDagsHubHome}, as discussed in \citet{lanubile2023training}, showcases the potential in this area.

The implications of these insights extend beyond the HF community, affecting the broader field of ML. They underscore the necessity for a paradigm shift in how ML models are conceptualized and maintained, potentially enhancing the efficiency, reliability, and overall effectiveness of ML development in community-driven platforms like HF.

\section{Threats to Validity}

In this section, we discuss the potential threats to the validity of our study and outline the mitigating actions we have taken to minimize these threats.

\textbf{Construct Validity:} Although we have employed comprehensive data collection and preprocessing methodologies, there is a possibility that the data may contain inaccuracies, inconsistencies, or missing values that could affect the results. This situation is exacerbated by the absence of standardized reporting for metadata on ML models. Moreover, to effectively measure constructs like popularity, maintenance, and evolution, we use relevant indicators and metrics. However, these might not fully capture the constructs' complexity, indicating a need for further research and refinement.

\textit{Mitigation:} We have implemented rigorous data cleaning and preprocessing procedures. We have also cross-validated the data obtained from the HF API with the \textit{HFCommunity} dataset to ensure consistency and completeness. For future studies, the implementation of model metadata extractors (e.g., \cite{tsay2020msr}) could be considered to enhance data quality further.

\textbf{Internal Validity:} Our classification of commits into corrective, perfective, and adaptive types is based on a neural network approach, which may introduce bias due to the training data or model architecture used.

\textit{Mitigation:} To mitigate this threat, we used a proven methodology from previous research and performed a validation check to ensure the accuracy and reliability of the commit classification. \citet{sarwar2020multi} reports a test accuracy of 89\%. We further manually analyzed 125 commit messages along its classifications to check the alignment of the results achieving ~86\% accuracy.

\textbf{External Validity:} Our study is based on data collected from the HF platform, which may limit the generalizability of our findings to other ML model platforms or communities. Additionally, our study focuses on the HF models as of November 6, 2023, and the findings may not be applicable to future developments on the platform.

\textit{Mitigation:} Our methodology is robust and replicable, designed to be applied to future datasets or similar platforms. We have provided a detailed methodology and a replication package to enable validation of our findings with new data, ensuring the broader applicability and relevance of our approach.

\textbf{Reliability:} Our study relies on a reproducible research methodology, where the data collection, preprocessing, and analysis procedures are clearly outlined. However, there is a possibility that changes in the HF API or \textit{HFCommunity} dataset structure could affect the reproducibility of our study.

\section{Conclusions}

This study presented a detailed examination of the evolution and maintenance of ML models on the HF platform, with a focus on two central research questions. The results offer a detailed understanding of the dynamics shaping model development and underscore the importance of systematic maintenance and incremental improvement for long-term model efficacy.

We observed that the HF community is not only expanding in terms of model quantity but also evolving through the adoption of new frameworks and tags, reflecting shifts in focus and innovation within the ML landscape, especially in the realm of generative AI. The analysis unveiled a vibrant ecosystem where certain models and tags gain prominence, indicative of the community's responsiveness to emerging trends and challenges in the field.

Moreover, our findings revealed that, while model development on HF encompasses both individual and collaborative efforts, there is a significant variance in the activity of model maintenance, as seen in the diverse distribution of commit patterns. Notably, perfective commits dominate, suggesting a continued focus on refining and optimizing models. The lifecycle of commits and the editing patterns of specific files further highlight different phases of model development, with an initial active development stage that transitions into stability and efficiency optimization in core components.

Additionally, we propose a framework for classifying models by their maintenance status, which could be instrumental for users in selecting models that align with their reliability and support requirements. Encouraging transparency in maintenance logs is essential in fostering a trust-based relationship between developers and the community. Moreover, the study highlights the unique maintenance dynamics of ML models on HF, diverging from traditional software paradigms with a focus on perfective maintenance, needing tools and methods tailored to ML's evolving needs.

For future work, developing advanced tools for automated and predictive maintenance to proactively address potential model issues is one crucial area. Investigating the social dynamics of model development can shed light on collaborative patterns, author roles, and best practices within the community. Extending this study to other ML platforms and comparing maintenance practices will provide insights into the broader ML development landscape.

In closing, this paper calls for concerted efforts towards analyzing the growth trajectory of ML model repositories while emphasizing the criticality of maintenance practices. We urge for enhanced transparency, structured maintenance frameworks, and community-wide standards that will propel the ML community towards greater heights of excellence and innovation.

\section*{ACKNOWLEDGMENTS}
This work is supported by the project TED2021-130923B-I00, funded by MCIN/AEI/10.13039/501100011033 and the European Union Next Generation EU/PRTR.

\bibliographystyle{IEEEtranN}

\newpage

\bibliography{References}

\end{document}